\documentclass[onecolumn]{svjour3}       
\usepackage{epsfig}
\usepackage{color}
\usepackage{amsmath}
\usepackage{amsfonts}
\usepackage{notes2bib}
\usepackage{graphicx}
\usepackage{amssymb}
\usepackage{subfig}

\DeclareMathOperator\erf{erf}

\newcommand{\brac}[1]{\left[#1\right]}
\newcommand{\curly}[1]{\left\{#1\right\}}
\newcommand{\para}[1]{\left(#1\right)}

\newcommand{\Rbar}{\ensuremath{\bar{\mathbf{R}}}}
\newcommand{\rb}{\ensuremath{{\mathbf{r}}}}

\journalname{Journal of Statistical Physics}

\begin{document}

\title{Response Theory for Static and Dynamic Solvation of Ionic and Dipolar Solutes in Water}

\titlerunning{Response Theory for Static and Dynamic Solvation}

\author{Renjie Zhao \and Richard C. Remsing \and John D. Weeks}

\institute{R. Zhao \at
		Institute for Physical Science and Technology
		University of Maryland, College Park, MD 20742
		\and
		R. C. Remsing \at
		Department of Chemistry and Chemical Biology,
		Rutgers University, Piscataway, NJ 08854 \\
		\email{rick.remsing@rutgers.edu}
		\and
		J. D. Weeks \at
		Institute for Physical Science and Technology
		 and Department of Chemistry and Biochemistry,
		 University of Maryland, College Park, MD 20742 \\
		 \email{jdw@umd.edu}
	}

\date{Received: date / Accepted: date}

\maketitle

\begin{abstract}
The response of polar solvents to ions and polar molecules dictates
many fundamental molecular processes. 
To understand such electrostatically-driven solvation processes, one ideally would probe the dielectric response
of a solvent to an idealized point test charge or dipole solute, as envisioned in classic continuum treatments of the problem.
However, this is difficult in simulations using standard atomically-detailed solvent models
with embedded point charges due to possible overlap with the test charge that
lead to singular interaction energies.
This problem is traditionally avoided for a realistic charged solute by
introducing an excluded volume core that shields its embedded point charge or dipole
from the charges in the solvent.
However, this core introduces additional molecular-scale perturbations of the solvent density that complicate the interpretation of solvent dielectric response.

In this work, we avoid these complications through the use of Gaussian-smoothed test charges and dipoles. 
Gaussian charges and dipoles can be readily inserted anywhere into an atomistic solvent model without encountering infinite energies.
If the Gaussian-smoothing is on the scale of molecular correlations in the solvent, both the thermodynamic and dynamic
solvation response is linear.
Using this observation, we construct accurate predictive theories for solvation free energies and solvation dynamics
for insertion of Gaussian charges and dipoles in polar solvents and demonstrate the accuracy of the theories
for a widely-used model of water.
Our results suggest that Gaussian test charge distributions can be used as an informative probe of dielectric response
in molecular models, and our theories can be used to analytically predict the largest component of solvation
free energies of charged and polar solutes.
%
\end{abstract}

\section{Introduction}
Solvation of charged and polar molecules is of fundamental importance to a wide range of processes in the chemical, materials, and biological sciences~\cite{BallReview}.
The thermodynamics of solvation determines whether or not dissolved molecules self-assemble and/or adsorb to interfaces,
for example.
The dynamic response of a solvent to chemical transformations, such as electron transfer reactions or electronic excitations,
can promote or hinder chemical reactions~\cite{Hynes:AnnRev:1985,gehlen1994dynamics,Pal:ChemRev:2004,Benjamin:2015bj,Orr_Ewing_2015,C7CS00331E}.
Thus, understanding the molecular-scale details underlying the thermodynamics and dynamics of solvation
is of critical importance to solution-phase chemistry and significant effort has focused
on developing theoretical models of these processes.
In this work, we revisit the development of simple theories of solvation thermodynamics and dynamics
in polar liquid solvents.
The central quantity in the thermodynamics of solvation is the solvation free energy, which quantifies
how favorable (or unfavorable) a solute is accommodated by a solvent. 
Classic models of ionic and polar solvation are rooted in dielectric continuum theory. 
In the simplest approximation, one imagines that the polar solvent is a uniform featureless dielectric medium with
dielectric constant $\varepsilon$. 
To probe the dielectric response, an approximation to the charged solute is then inserted into the uniform dielectric solvent. However one cannot simply insert an idealized point test charge or dipole directly into the dielectric because
 the insertion energy tends to infinity due to the singularity in the Coulomb potential as $r\rightarrow0$. 
To avoid this issue, classic approaches introduce the concept of an idealized solute core,
a (often spherical) boundary inside of which there is no dielectric solvent, while outside there is still the unperturbed uniform dielectric.
By introducing this boundary condition, the energy and free energy are finite and can be evaluated at the level of dielectric continuum theory,
although the physical interpretation of this boundary is not well understood.
The most well-known versions of these dielectric continuum theory-based solvation models
are the Born and Bell models~\cite{Born,bell1931electrostatic}, achieved by placing a point charge (monopole) or point dipole
inside a spherical solute core, and this class of models can be easily generalized to insert any multipole (or set of multipoles) inside the cavity. 
While the Born and Bell models can reproduce the correct magnitude of solvation free energies with reasonable core sizes (even when neglecting the free energy
required to create the solute core), the absence of solvent structuring at the solute core boundary can lead
to qualitative inaccuracies when comparing to experiment.

These inaccuracies arise from non-linear solvent response to a physically realistic solute model.
In water, for example, the Born model predicts ionic solvation free energies that are symmetric with respect to the charge
of the ion, but in reality these free energies are asymmetric and non-linear with respect to ionic charge
for ions of the same size.
This non-linear response is due to the asymmetric interfacial structure induced by the solute core --- water dipoles
preferentially orient toward the core even when it is not charged --- as well as an inherent non-linear response
associated with inserting a rapidly-varying, harshly repulsive potential into the solvent.
Previous work has shown that an exact reordering of the solvation process can isolate the dominant
and linear electrostatic component of the solvation free energy, relegating the non-linear processes involving
solute cores to smaller but conceptually important contributions~\cite{Remsing:JPCB:2016}.
This reordering involves three steps: (1) insert a \emph{Gaussian} solute charge distribution into the solvent,
(2) insert the physical uncharged solute core, and (3) shrink the Gaussian charge distribution to the point multipolar distribution of the solute.
Replacing the point multipoles of the solute by a Gaussian smeared analog in the first step removes the singularity as $r\rightarrow0$,
accomplishing the same task as creating the solute core in the Born and Bell models.
Choosing the width of the Gaussian to be on the order of molecular correlations, roughly 3-5~\AA \ in water,
ensures that the first step in the process is linear and can be evaluated analytically. 
This first step encompasses roughly eighty percent of the total charging free energy~\cite{Remsing:JPCB:2016}.
Moreover, insertion of a Gaussian charge distribution into a charged or polar solvent
enables the isolation of the dielectric response of the solvent independent of any detailed solute core effects.
Thus, Gaussian charge distributions also provide a useful and sensitive probe of solvent dielectric response,
in direct analogy to what one would like to do with point test charges in classic electrostatics contexts.
Here, we extend this work in two important ways.
First, we develop a theory for the solvation of Gaussian dipoles in polar solvents in analogy to the Bell model.
The use of Gaussian charge distributions ensures the quantitative accuracy of this model for describing solvation in
atomically-detailed solvent models, and we validate this through comparison of our theory to results from
molecular dynamics (MD) simulations.
We then extend the Gaussian solvation models to describe solvation dynamics.
Solvation dynamics refers to the time-dependent response of a solvent to changes in the nature of a solute~\cite{Bagchi:AnnRev:1989,Carter:JCP:1991,Maroncelli:Nature,Stratt:JPC:1996,Castner_1988,song1996gaussian,Geissler:JCP:2000,Benjamin:2006aa,Thompson:2011ax}.
Traditionally, solvation dynamics focuses on the solvent response to instantaneous changes in a solute
charge distribution, physically realized by electronic excitations, for example.
Such charge distributions can be accurately approximated by point charges or dipoles within physical cores.
Traditional approaches to solvation dynamics utilize dielectric theory to interpret results and
draw conclusions about the nature of dynamic dielectric response.
However, as noted above in the context of solvation thermodynamics, the presence of excluded volume cores complicates
such interpretations and models because these cores induce nonlinear responses that are not well-described
by dielectric continuum theories. 
By using Gaussian charge and dipole distributions, we are able to probe an intrinsic component of the dielectric response of a model solvent
unobscured by the presence of excluded volume cores.
Isolating the dynamic dielectric response in this manner enables the development and testing of molecular theories for
dielectric solvation dynamics without implicit assumptions of the nature of the solvent response to solute cores.
Such theories can then be extended to systems with cores in a systematic manner.

Indeed, we demonstrate that the solvent responses to molecular scale ionic and dipolar Gaussian charge distributions are linear.
Moreover, we find that the collective dynamics of the solvent response are well-described by a linear response theory
involving the frequency-dependent dielectric constant, $\varepsilon(\omega)$. 
Our work highlights the utility of Gaussian charges as a modeling tool to understand the intrinsic response of a solvent to electrostatic perturbations.
Extensions of our work to polarizable and \emph{ab initio} models of water will further test
the accuracy of our theories~\cite{Remsing:2018aa,Duignan2017,laury2015revised,Lemkul:2016aa,Cisneros:2016aa,Duignan2017a,SCANWater,Duignan2019}.
We anticipate that such extensions could serve as useful tools to compare the dielectric response
of various polarizable and/or \emph{ab initio} models on an equal footing.

\section{Charging Free Energy of Ionic Solutes}
Before discussing the solvation thermodynamics of dipolar solutes in detail, we first summarize
previous results~\cite{Remsing:JPCB:2016} regarding the solvation free energy of two idealized models of charged ionic solutes:
a point charge located at the center of a hard sphere excluded volume core (the Born model)
and a Gaussian charge distribution.
In both cases, we are concerned with the free energy change upon introduction of a solute-solvent Coulomb interaction,
where the ionic solute has a net charge of $\lambda Q$, such that $\lambda$ is a (linear) coupling parameter that we can
use to `turn on' the charge of the solute.
This solute-solvent electrostatic interaction energy is given by
\begin{equation}
\Psi_\lambda(\Rbar)=\int d\rb \int d\rb' \frac{\rho^Q_\lambda(\rb')\rho^q(\rb;\Rbar)}{\left| \rb-\rb'\right|},
\label{eq:psi}
\end{equation}
where
\begin{equation}
\rho^q(\rb;\Rbar)=\sum_{i=1}^{N_C} q_i \delta (\rb-\rb_i(\Rbar))
\end{equation}
is the solvent charge density in configuration $\Rbar$, the solvent is composed of $N_C$ charged sites located at positions $\rb_i(\Rbar)$,
and $\rho^Q_\lambda(\rb)$ is the solute charge density.
For a point charge located at the origin, $\rho^Q_\lambda(\rb)=\lambda Q \delta(\rb)$,
while for a Gaussian charge distribution of width $l$ and magnitude $Q$, $\rho^Q_\lambda(\rb)=\lambda Q\rho_G(r;l)$,
where
\begin{equation}
\rho_G(r;l)=\frac{1}{l^3\pi^{3/2}}e^{-r^2/l^2}.
\end{equation}
In the limit $l\rightarrow0$, the Gaussian distribution approaches the point charge distribution. 
As shown in~\cite{Remsing:JPCB:2016}, the charging free energy of turning on the solute charge, $\Delta G^c(Q)$,
can be exactly obtained through coupling parameter integration as
\begin{equation}
\Delta G^c(Q) = \int_0^1 d\lambda \int d\rb \int d\rb' \frac{\rho^Q(\rb') \rho^q_\lambda(\rb)}{\left|\rb-\rb'\right|},
\label{eq:gfe}
\end{equation}
where $\rho^q_\lambda(\rb)=\left<\rho^q(\rb;\Rbar)\right>_\lambda$ is the ensemble average of the solvent charge density
in state $\lambda$.
Typical dielectric continuum theory-based approaches assume that the average solvent charge density in the absence of a solute
charge is zero, $\rho^q_0(\rb)=0$.
This is true for the case of Gaussian charge insertion into a bulk solvent, but not for the situation where a solute excluded volume core
exists prior to turning on the solute charge; see previous work for more details~\cite{Remsing:JPCL:2014,Remsing:JPCB:2016,Remsing:2018aa,Remsing2019,BeckReview,HummerIonHydration,Garde,GeisslerReview,BallReview,Duignan2017,Duignan_2018,Beck_2018}.
However, working within this assumption is important to connect to classic solvation models, e.g. the Born model.
For slowly varying solute charge distributions and within linear response theory, we have shown in \cite{Remsing:JPCB:2016} that the induced solvent charge density can be linearly and locally related to the overlapping solute charge density, effectively canceling all but a fraction $1/\epsilon$ of
the solute charge:
\begin{equation}
\rho^q_\lambda(r)=-\left(1-\frac{1}{\varepsilon}\right) \lambda \rho^Q(r).
\label{eq:overlap}
\end{equation}
For the idealized linear dielectric solvent of the Born model, Eq. (\ref{eq:overlap}) is assumed to hold exactly for any overlapping solute charge density. In particular for the Born model of ionic solvation, we can use Gauss's law to replace the central point charge by an overlapping charge distribution at the Born radius, $R_{\rm B}$, of the form $\rho^Q(r)=Q \delta(r-R_{\rm B})/4\pi r^2$ for $r\ge R_{\rm B}$ and 0 for $r<R_{\rm B}$, and the resulting charging free energy from Eqs.\ \ref{eq:gfe} and \ref{eq:overlap} is
\begin{equation}
\Delta G^c(Q) = - \frac{Q^2}{2R_{\rm B}}\left( 1-\frac{1}{\varepsilon}\right).
\end{equation}
This is the Born model of ion solvation~\cite{Born,Remsing:JPCB:2016}. 
The Born model can reproduce the general magnitude of ionic solvation free energies, but its neglect of structural
and electrostatic effects that
arise from the uncharged excluded volume core results in qualitative and quantitative inaccuracies. 
Experimentally-determined and simulated ionic solvation free energies
for ions with the same size cores are asymmetric with respect to the sign of the ion charge,
while the Born model predicts that $\Delta G^c(Q)=\Delta G^c(-Q)$.
This asymmetry results from a non-zero $\rho_0^q(r)$; the introduction of an uncharged ionic core induces an electrostatic
response in the solvent that must be taken into account.
Additional non-linear responses also arise due to the presence of this harshly repulsive core~\cite{Remsing:JPCB:2016}.
For a Gaussian solute charge distribution, defined above, the resulting free energy is
\begin{equation}
\Delta G^c(Q;l)=-\frac{Q^2}{l\sqrt{2\pi}}\left( 1-\frac{1}{\varepsilon}\right).
\label{eq:gcfe}
\end{equation}
The two model free energies become equivalent in magnitude
for a Gaussian of width $l_{\rm B}$, where $R_{\rm B}=l_{\rm B}\sqrt{\pi/2}$.
Unlike the charging of a hard sphere, simulation results indicate that
the charging of a Gaussian charge distribution is symmetric with respect to the sign of the charge
and follows linear response theory for values of $l$ that are on the scale of charge-charge correlations in the solvent (or larger). 
Thus, the dielectric continuum theory-like linear response theory for $\Delta G^c(Q)$ yields accurate quantitative predictions. 

In the derivation of the above free energies, as well as those given below, we have, for simplicity, assumed zero pre-existing solvent charge density arising from hypothetical and/or structural boundaries, \emph{i.e.} we consider an infinite solvent~\cite{Remsing2019}.
These boundary terms contribute constants of the form $Q\phi$, where $\phi$ is a constant potential originating from
nonuniform solvent charge densities at electrostatic boundaries,
and only arise for non-neutral systems, e.g. single-ion solvation free energies.
By working in the infinite solvent limit, we omit these terms without loss of generality,
and they can be readily included using the treatment given elsewhere~\cite{Remsing2019}.

\section{Charging Free Energy of Dipolar Solutes}
As demonstrated above, the solvation thermodynamic properties predicted by dielectric continuum treatments differ depending on the precise models adopted.
In contrast to the Born model, which inherently ignores nonlinear responses associated with the presence of an excluded volume solute core, Gaussian test charges provide a testing ground for elucidating the microscopic origins of linear and nonlinear responses in ion solvation. 
To extend this viewpoint to the charging free energy of dipolar solutes,
we explore the solvation of Gaussian-smoothed dipoles within the framework of linear response theory
and compare our results to the classic Bell model, the analog of the Born model for dipolar solutes.

\subsection{Bell Model Revisited}
Bell introduced the dipolar generalization of the Born model by idealizing a dipolar solute as a point dipole fixed at the center of a spherical cavity of radius $R$~\cite{bell1931electrostatic}, and we first revisit this concept before describing its Gaussian analog.
We consider a charging process in which the solute charge is a partially charged point dipole with electrostatic potential
$v_\lambda^\mu\left(\mathbf{r}\right)=\lambda v^\mu(\mathbf{r})$.
Here $v^\mu(\mathbf{r})$ is the electrostatic potential of a point dipole with a dipole moment along the $z$-axis of magnitude $\mu$, 
\begin{equation}
v^\mu\left(\mathbf{r}\right)=\frac{\mu\cos{\varphi}}{r^2},
\label{eq:pointDipole}
\end{equation}
and $\lambda$ is a linear coupling parameter varying between zero and unity.
As with Gaussian charges described above, the charging component of the solvation free energy can be obtained from a coupling parameter integration, such that Eq.~\ref{eq:gfe} reduces to
\begin{equation}
\Delta G^c\left(\mu\right)=\int{d\mathbf{r}v^\mu(\mathbf{r})}\int_{0}^{1}{d\lambda\rho_\lambda^q(\mathbf{r})}.\label{eq:charging}
\end{equation}
%

The overall electrostatic potential due to both the solute charge and the induced solvent charge density can be obtained by solving Poisson's equation with the relevant boundary condition, which gives
\begin{equation}
  v_\lambda^{\rm Bell}(r,\varphi) =
    \begin{cases}
      \displaystyle{\frac{3}{2\varepsilon+1}\frac{\lambda\mu\cos{\varphi}}{r^2}} & \text{$r>R$}\\
      \displaystyle{\frac{\lambda\mu\cos{\varphi}}{r^2}-\frac{2\left(\varepsilon-1\right)}{2\varepsilon+1}\frac{\lambda\mu r\cos{\varphi}}{R^3}} & \text{$r<R$ ,}
    \end{cases} 
\end{equation}
where $\varepsilon$ is the dielectric constant of the solvent and we have assumed $\vec{\mu}=\mu \hat{z}$ without
loss of generality; the dipole is aligned with the $z$-axis.
The induced solvent charge density $\rho_\lambda^q(\mathbf{r})$ distributed over the surface of the cavity is computed by
\begin{equation}
4\pi\rho_\lambda^q\left(\mathbf{r}\right)=\left.-\frac{\partial v_\lambda^{Bell}}{\partial r}\right|_{r=R+\epsilon}+\left.\frac{\partial v_\lambda^{Bell}}{\partial r}\right|_{r=R-\epsilon}
\end{equation}
and we obtain
\begin{equation}
\rho_\lambda^q\left(\mathbf{r}\right)=-\frac{3\lambda\mu\delta(r-R)\cos{\varphi}}{2\pi r^3}\frac{\varepsilon-1}{2\varepsilon+1}.\label{eq:surfdens}
\end{equation}
Since the induced charge density of the Bell model depends linearly on the coupling parameter $\lambda$, we can integrate out $\lambda$ and Eq.~\ref{eq:charging} reduces to
\begin{equation}
\Delta G^c\left(\mu\right)=\frac{1}{2}\int{d\mathbf{r}v^\mu\left(\mathbf{r}\right)\rho^q\left(\mathbf{r}\right)},\label{eq:reducedcharging}
\end{equation}
where $\rho^q\left(\mathbf{r}\right)=\rho_{\lambda=1}^q\left(\mathbf{r}\right)$ is the induced charge density of the fully coupled solute-solvent system.

By inserting Eq.~\ref{eq:surfdens} and Eq.~\ref{eq:pointDipole} into Eq.~\ref{eq:reducedcharging}, we obtain the charging free energy of the Bell model
\begin{align}
\Delta G^c\left(\mu\right)
&=  -\frac{3\mu^2}{4\pi}\frac{\varepsilon-1}{2\varepsilon+1}\int_{0}^{2\pi}d\theta\int_{0}^{\pi}{d\varphi\cos^2\varphi\sin{\varphi}}\int_{R}^{\infty}{dr\frac{\delta\left(r-R\right)}{r^3}}  \\
&=  -\frac{\varepsilon-1}{2\varepsilon+1}\frac{\mu^2}{R^3}.
\label{eq:E_Bell}
\end{align}

Note that the solvent charge density of the Bell model is nonzero only on the surface of the cavity and is therefore nonlocal as induced from the point dipole, similar to the solvent charge density of the Born model.
This discontinuity in the charge density is a consequence of applying macroscopic electrostatic treatments to an atomically sharp boundary. Moreover, the complicated $\varepsilon$ dependence of the charging free energy in Eq.~\ref{eq:E_Bell} is directly inherited from the $\varepsilon$ dependence of the solvent charge density in Eq.~\ref{eq:surfdens}.

\subsection{Gaussian-Smoothed Dipole Model}

Slowly-varying charge distributions, such as Gaussian test charges, generate spatially slowly varying electric fields that induce local and linear dielectric responses even in realistic solvents.
By careful construction, such slowly-varying charge distributions can be used to account for long-ranged electrostatic interactions in more complex systems of interest and its charging free energy constitutes the dominant component of the total solvation free energy of a dipolar molecule, for example.

\subsubsection{Constructing a Gaussian-smoothed dipole}
\begin{figure*}[tb]
\begin{center}
\includegraphics[width=0.98\textwidth]{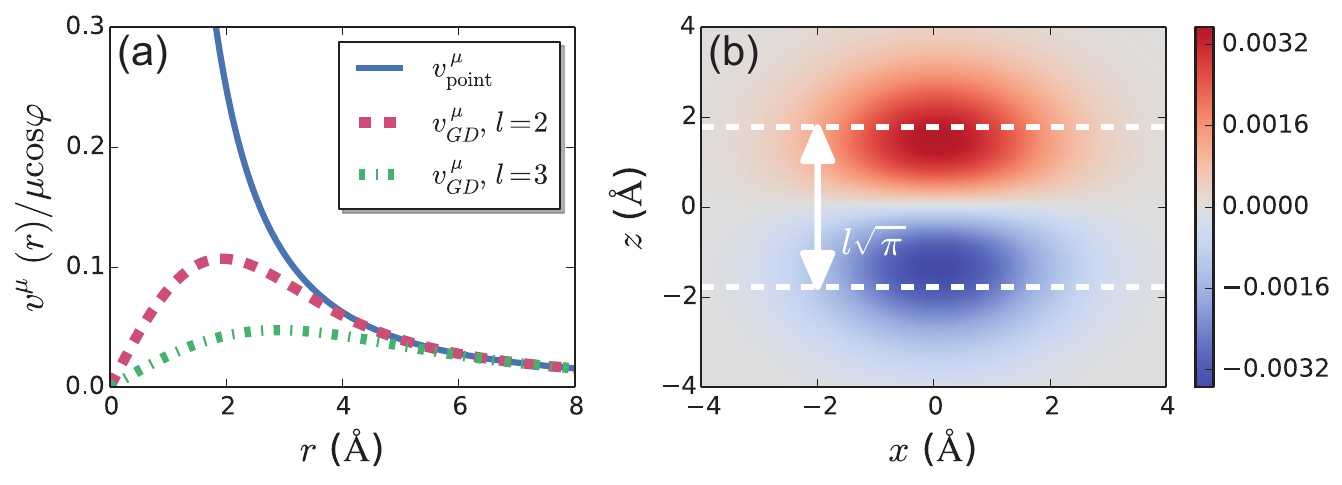}
\end{center}
\caption[dipole radial potential]
{
(a) Comparison between the radial part of the point dipole potential $v_{point}$ and the radial part of the Gaussian dipole potential $v_{GD}$ for two different values of smoothing length $l$. In the limit as $l \to 0$, $v_{GD}$ reduces to $v_{point}$. With greater $l$, $v_{GD}$ becomes more slowly varying.
(b) Charge density of the Gaussian dipole (Eq.~\ref{eq:density}) scaled by the magnitude of its dipole, $\rho^\mu_{\rm GD}(\rb)/\mu$,
in the $xz$-plane for fixed $y=l$, with $l=2$~\AA. 
White dashed lines indicate $z=\pm\sqrt{\pi}l/2$, highlighting the upper-bound on the effective distance between the two charges
comprising the dipole given by Eq.~\ref{eq:dbar}.
}
\label{fig:potential}
\end{figure*}

The Gaussian-smoothed dipole, generalized from the Gaussian test charge,
is a natural candidate for the dipole solute with slowly-varying charge distribution.
We can construct a Gaussian-smoothed dipole from two oppositely charged Gaussian charges of the same width, $l$.
The electrostatic potential due to a single unit Gaussian charge distribution with a smoothing length $l$ is
\begin{equation}
v_G\left(r\right)=\frac{\erf{(r/l)}}{r}.
\end{equation}
A Gaussian-smoothed dipole can then be realized by placing
a Gaussian test charge of magnitude $Q$ centered at the origin and another Gaussian test charge of magnitude $-Q$ at an infinitesimal distance $-d\mathbf{r}$ from the origin.
The potentials arising from each individual charge distribution are $Qv_G(\mathbf{r})$ and $-Qv_G(\mathbf{r}+d\mathbf{r})$, respectively.
The potential of the composite Gaussian-smoothed dipole is given by
\begin{align}
v_{GD}\left(\mathbf{r}\right)
&=-Q\{v_G\left(\mathbf{r}+d\mathbf{r}\right)-v_G\left(\mathbf{r}\right)\}  \\
&=  -Qd\mathbf{r}\cdot\nabla v_G\left(\mathbf{r}\right).
\end{align}
Similar to the limiting case where a point dipole is traditionally constructed from two point charges~\cite{Zangwill},
the product $Qd\mathbf{r}$ defines the dipole moment $\boldsymbol{\mu}$ and is kept finite.
We may therefore write the potential of the Gaussian-smoothed dipole with width $l$ as
\begin{equation}
v_{GD}^\mu\left(\mathbf{r}\right)=-\boldsymbol{\mu}\cdot\nabla\frac{\erf{(r/l)}}{r}
\label{eq:Pt_GD}
\end{equation}
Assuming, without loss of generality, that $\boldsymbol{\mu}=\mu\hat{z}$, Eq.~\ref{eq:Pt_GD} can be expressed as
\begin{equation}
v_{GD}^\mu\left(r,\varphi\right)=\mu\cos{\varphi}\left[\frac{\erf{(r/l)}}{r^2}-\frac{2e^{-r^2/l^2}}{r l \sqrt\pi}\right],
\label{eq:Pt_GD_sph}
\end{equation}
which asymptotically approaches the well-known point dipole potential $v_{\rm point}(r,\varphi)=\displaystyle{\frac{\mu\cos{\varphi}}{r^2}}$ on length scales greater than $l$.
As illustrated in Figure~\ref{fig:potential}a, the potential arising from the Gaussian dipole is slowly varying with sufficiently large $l$
and removes the singularity in the point dipole potential without requiring the shielding, excluded volume core employed in the Bell model.

\subsubsection{Relation to a molecular dipole}

A traditional approach to modeling a molecular dipole is to place two point charges of magnitude $|Q|$
and opposite sign a distance $d$ apart~\cite{Zangwill}. 
The dipole moment of this model is readily given by $\mu=Qd$,
such that the distance $d$ is a lengthscale inherent to the dipole.
A similar lengthscale emerges from our Gaussian smoothed dipole moment.

The charge density distribution of the Gaussian dipole is obtained by applying Poisson's equation to Eq.~\ref{eq:Pt_GD_sph}, 
\begin{equation}
\rho_{GD}^\mu\left(\mathbf{r}\right)=\mu\cos{\varphi}\frac{2re^{-r^2/l^2}}{\pi^{3/2}l^5}.
\label{eq:density}
\end{equation}
By construction, the centers of the two Gaussian charges are infinitesimally separated,
but it can be observed in the contour of the charge density that positive and negative charges are concentrated at a finite molecular distance from each other, Fig.~\ref{fig:potential}b.
To connect to the point charge dipole model,
we define the magnitude of net positive (negative) charge in the upper (lower) half space by
\begin{align}
Q_{net}
&=\int_{upper}{d\mathbf{r}\rho_{GD}^\mu\left(\mathbf{r}\right)} = \frac{\mu}{l\sqrt\pi}.
\label{eq:qnet}
\end{align}
The point charge dipole yields $Q=\mu/d$, and comparison of this expression with Eq.~\ref{eq:qnet}
defines an effective separation $\bar{d}$.
However, there will be partial cancelation of charges due to the smearing of the Gaussian charge distributions,
i.e. some negative charge leaks into the upper half space and vice versa,
such that 
\begin{equation}
Q_{net}\lesssim Q.
\end{equation}
This enables us to place an upper bound on the effective separation between charges in the Gaussian dipole:
\begin{equation}
\bar{d}\lesssim l\sqrt\pi.
\label{eq:dbar}
\end{equation}
This bound is shown in Fig.~\ref{fig:potential}b and indeed gives a reasonable approximation to the distance between
the two lobes of the dipole distribution.
This result enables comparison between the Gaussian-smoothed dipole and traditional dipolar solute models
composed of two spatially-separated point charges.

\subsubsection{Solvation free energy of a Gaussian-smoothed dipole}

\begin{figure*}[tb]
\begin{center}
\includegraphics[width=0.75\textwidth]{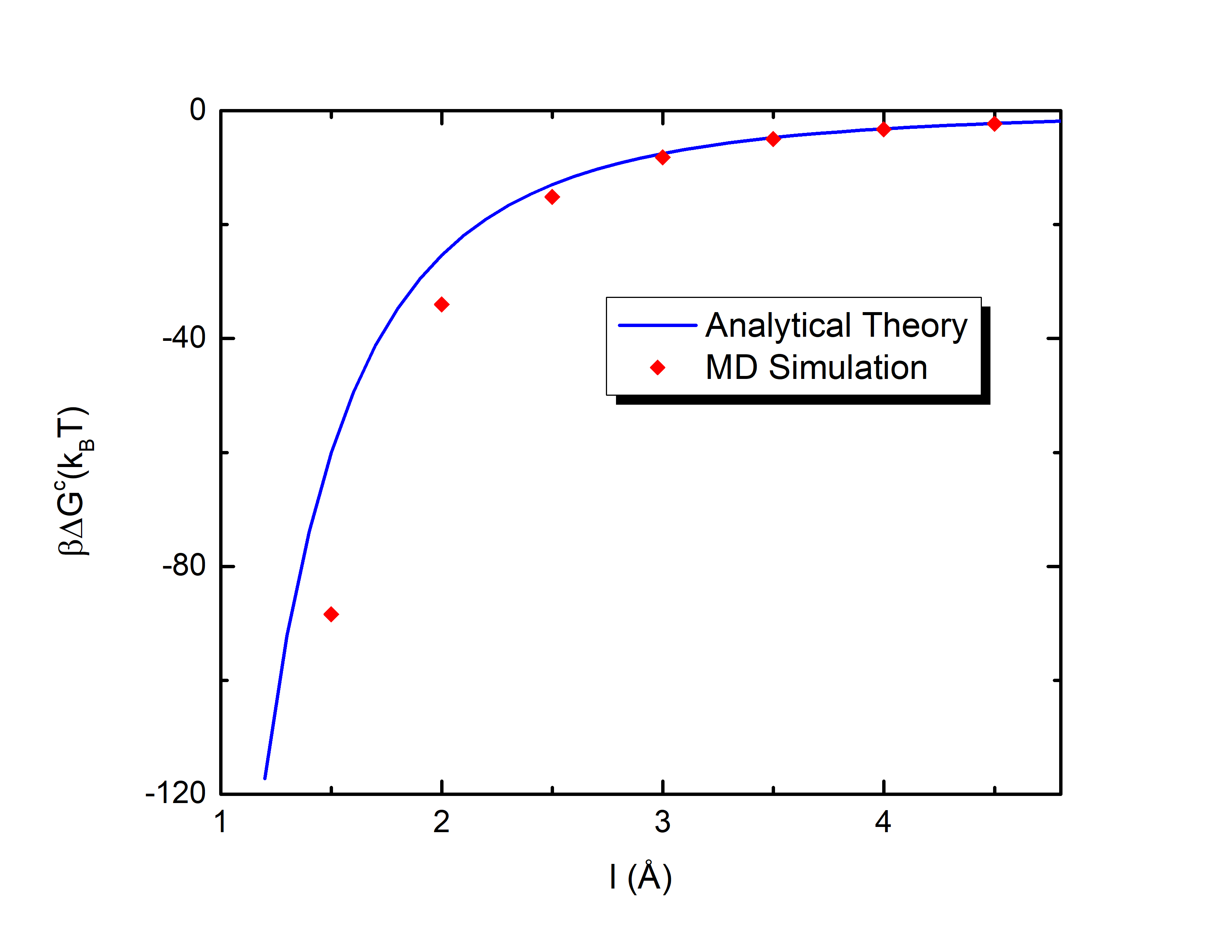}
\end{center}
\caption[free energy]
{
Charging free energy of a Gaussian dipole with $\mu=8$~D and various values of smoothing length $l$.
Data points correspond to simulation results and error bars are smaller than the symbol size.
The solid line is the prediction of Eq.~\ref{eq:G_freeEnergy}.
}
\label{fig:energy}
\end{figure*}

In order to determine the solvation free energy of the Gaussian dipole, we note
that its slowly varying nature, Eq.~\ref{eq:density},
enables the use of approximations that exploit the local and linear response of the solvent to the solute.
We consider the charging process of a Gaussian dipole in the dielectric continuum
by coupling its potential to a parameter $\lambda$ varying from 0 to 1, such that
the solute charge density is $\rho_{GD}^{\lambda\mu}\left(\mathbf{r}\right)=\lambda\rho_{GD}^\mu\left(\mathbf{r}\right)$.
Linear response theory approximates the solvent charge density by
\begin{equation}
\rho_\lambda^q\left(\mathbf{r}\right)\approx-\left(1-\frac{1}{\varepsilon}\right)\rho_{GD}^{\lambda\mu}\left(\mathbf{r}\right),
\label{eq:linear}
\end{equation}
such that the dielectric continuum screens the solute charges by a fraction $\left(1-\displaystyle{{1}/{\varepsilon}}\right)$.
Again, assuming no pre-existing solvent charge density in this model, the charging free energy can be computed with a form analogous to Eq.~\ref{eq:charging},
\begin{equation}
\Delta G^c\left(\mu\right)=\int{d\mathbf{r}v_{GD}^\mu\left(\mathbf{r}\right)}\int_{0}^{1}{d\lambda\rho_\lambda^q\left(\mathbf{r}\right)}.
\label{eq:Gaussian_Charging}
\end{equation}
By inserting Eq.~\ref{eq:Pt_GD} and Eq.~\ref{eq:linear} into Eq.~\ref{eq:Gaussian_Charging}, we obtain
\begin{equation}
\Delta G^c\left(\mu\right)=-\left(1-\frac{1}{\varepsilon}\right)\frac{\mu^2}{3\sqrt{2\pi}l^3}.
\label{eq:G_freeEnergy}
\end{equation}
In the regime where linear response theory holds,
e.g. for a Gaussian dipole with moment $8$~D, this corresponds to a width of $l\ge3$~\AA,
the free energy computed from molecular dynamics (MD) simulations agrees well with the above continuum theory,
as shown in Fig.~\ref{fig:energy}.
The magnitude of the dipole moment, $8$~D, is roughly the change in the dipole moment
upon electronic excitation of coumarin 153, one of the most widely used probe molecules in
experimental solvation dynamics studies~\cite{Stratt:JPC:1996}.

To compare the charging free energy of the Gaussian dipole with that of the Bell model Eq.~\ref{eq:E_Bell}, we write Eq.~\ref{eq:G_freeEnergy} in an alternative form
\begin{equation}
\Delta G^c\left(\mu\right)=-\left(\frac{\varepsilon-1}{2\varepsilon}\right) \frac{\mu^2}{\left[\left({9\pi}/{2}\right)^{1/6}l\right]^3}.
\end{equation}
The solvation free energies of the two models are first different by a factor of $\frac{2\varepsilon+1}{2\varepsilon}$.
This factor approaches $\frac{3}{2}$ in the limit of vacuum, while it approaches unity for solvents with large dielectric constants such as water.
The origin of $\frac{2\varepsilon+1}{2\varepsilon}$ is the different dielectric continuum treatments applied to the two models.
In the Gaussian dipole model,
the linear response theory approximation that we adopted is consistent with dielectric continuum screening arguments,
which leads to the same $\varepsilon$ dependence as models of ion solvation.
This is also expected according to the superposition principle of electrostatics, given that the Gaussian dipole is made of two Gaussian charges.
In contrast, as mentioned above, the $\varepsilon$-dependence of the Bell model arises from Eq.~\ref{eq:surfdens},
the singular charge distribution of the solvent.
Furthermore, with different orders of multipole moments,
Born- and Bell-like models will generally exhibit different $\varepsilon$-dependences of the free energy due to the
boundary condition imposed by the excluded volume.
In contrast, the Gaussian models do not exhibit such a multipole-dependent $\varepsilon$-dependence.

In cases where the difference due to this $\varepsilon$-dependent prefactor is negligible,
the free energies of the Bell and the Gaussian dipole models become equivalent
when the hard-core radius of the Bell model is equal to
\begin{equation}
R=l\left(\frac{9\pi}{2}\right)^{{\frac{1}{6}}}.
\label{eq:bellR}
\end{equation}

The solvation free energy of a Gaussian dipole is an important component of the total solvation free energy
of a molecularly-detailed dipolar molecule.
This can be seen through a judicious rewriting of the solvation free energy~\cite{Remsing:JPCB:2016}, wherein
a Gaussian dipole is first inserted into the solvent, then the molecular core is inserted, and finally
the width of the Gaussian is decreased to the point dipole limit or morphed into a more detailed molecular charge distribution.
However, unlike a Gaussian charge distribution, a Gaussian dipole can be completely screened by a dipolar solvent
like water, and this complete screening results in a lower solvation free energy.
Despite this point, the free energy of this first step of inserting the Gaussian dipole is expected to
be a significant fraction of the electrostatic contribution to the solvation free energy, 
and Eq.~\ref{eq:G_freeEnergy} provides an accurate analytic expression for this quantity.
This is especially true for small solute molecules, where the free energy of inserting an excluded volume core
is less than or comparable to the free energies in Fig.~\ref{fig:energy} in the linear response regime~\cite{Remsing:JPCB:2016}.
For large dipolar solutes, such as charge-neutral proteins, the Gaussian dipole charging free energy will be an important
contribution to the total solvation free energy, although
the relative partitioning of the free energy components will depend sensitively on $l$,
with Eq.~\ref{eq:bellR} yielding $l\sim 1$~nm for a small protein~\cite{Remsing_2018}.

%

\section{Linear Response Theory for Dynamic Response to Gaussian Charges}

Solvation dynamics concerns the time evolution of the solvent dielectric response to a time-dependent change in the solute charge distribution,
for example, through creation of an ion or a change in solute dipole moment through photoexcitation~\cite{Bagchi:AnnRev:1989,Stratt:JPC:1996,Benjamin:2006aa,Thompson:2011ax}.
Classic Born-like models have therefore been extended to nonequilibrium scenarios to describe the dynamics of ionic and dipolar solvation, enabling successful predictions for the time scales on which solvation free energy decays.
Here, we focus on extending the Gaussian-smoothed solute models into this phenomenological nonequilibrium theory,
and our approach follows the framework of linear irreversible processes~\cite{toda1991statistical}.

We consider turning on the solute charge distribution instantaneously in an equilibrated bulk solvent.
This introduction of the solute charge distribution, $\rho^Q_G(\rb,t)$, defines $t=0$. 
We then focus on how the solvent dynamically responds to the the instantaneous introduction
of the solute charge and evolves to a new equilibrium state. 
The dynamical response of the solvent is manifest in its charge density $\rho^q\left(\mathbf{r},t\right)$.
%
We can estimate the time-dependent induced solvent charge density by
\begin{equation}
\rho^q\left(\mathbf{r},t\right)=-\left(1-\frac{1}{\varepsilon_h}\right)\rho_G^Q\left(\mathbf{r},t\right)+\int_{-\infty}^{t}{dt^\prime}\Phi\left(t-t^\prime\right)\rho_G^Q\left(\mathbf{r},t^\prime\right).
\label{eq:charge_DynmicalResp}
\end{equation}
The first term in Eq.~\ref{eq:charge_DynmicalResp} represents an effectively instantaneous response of the solvent
to the introduction of the solute charge.
Physically, this instantaneous response results from a coarse-graining of high frequency modes of
the solvent polarization response, such that the parameter $\varepsilon_h$ contains the effects
of these high frequency modes.
The connection of $\varepsilon_h$ to the frequency-dependent dielectric constant, $\varepsilon(\omega)$,
and its precise value will be discussed in more detail below.
%
%
The second term describes non-Markovian effects due to the time lag of the polarization of the solvent in response to the time-dependent solute charge distribution, where $\Phi\left(t-t^\prime\right)$ is the response function of the considered system, which will be determined below.
Because the solute distribution will be turned on instantaneously at $t=0$, the integrand is non-zero only
for $t'>0$, but we use this form for generality.
Given that $\rho^{eq}=0$ due to neutrality, Eq.~\ref{eq:charge_DynmicalResp} follows directly from the standard framework of a linear irreversible process.

For mathematical convenience, we change the integration variable from $t^\prime$ to $s=t-t^\prime$ and obtain
\begin{equation}
\rho^q\left(\mathbf{r},t\right)=-\left(1-\frac{1}{\varepsilon_h}\right)\rho_G^Q\left(\mathbf{r},t\right)+\int_{0}^{\infty}{ds\ \Phi\left(s\right)\rho_G^Q\left(\mathbf{r},t-s\right)}.
\label{eq:charge_DynmicalResp2}
\end{equation}
Taking the Laplace (one sided Fourier) transform of Eq.~\ref{eq:charge_DynmicalResp2} yields
\begin{equation}
\rho^q(\rb,\omega)=-\para{1-\frac{1}{\varepsilon_h}}\rho^Q_G(\rb,\omega) + \Phi(\omega) \rho^Q_G(\rb,\omega),
\label{eq:charge_DynmicalResp3}
\end{equation}
where $\Phi(\omega)=\int_0^\infty dt \exp(-i\omega t) \Phi(t)$.
Equation~\ref{eq:charge_DynmicalResp3} describes the response of the solvent to a charge distribution harmonically oscillating with frequency $\omega$.

Alternatively, this response can be analyzed by generalizing Eq.~\ref{eq:linear} to the dynamical domain as
\begin{equation}
\rho^q\left(\mathbf{r},\omega\right)\approx-\left(1-\frac{1}{\varepsilon\left(\omega\right)}\right)\rho_G^Q\left(\mathbf{r},\omega\right),
\label{eq:linear_omega}
\end{equation}
where $\varepsilon\left(\omega\right)$ is the frequency-dependent dielectric constant.
Comparing Eq.~\ref{eq:charge_DynmicalResp3} and Eq.~\ref{eq:linear_omega} yields an expression for the response function
\begin{equation}
\Phi(\omega)=\frac{1}{\varepsilon\left(\omega\right)}-\frac{1}{\varepsilon_h}.
\label{eq:compare}
\end{equation}
This expression for the response function can be used to described the time-dependence of the
induced solvent charge density according to
\begin{equation}
\rho^q(\rb,t)=-\rho^Q_G(\rb,t) + \int_0^\infty ds \rho_G^Q(\rb,t-s) \mathcal{L}^{-1}\curly{\frac{1}{\varepsilon(\omega)}},
\label{eq:general}
\end{equation}
where $\mathcal{L}^{-1}\curly{f(\omega)}$ indicates the inverse Laplace transform of $f(\omega)$.
Equation~\ref{eq:general} is the linear response theory estimate for $\rho^q(\rb,t)$ to a time-dependent
solute charge density, which we expect to be valid for a spatially slowly-varying Gaussian charge distribution.
%
%

%
In this work, we consider turning on a solute charge density instantaneously at $t=0$, such that the
time-dependent solute charge density is
\begin{equation}
\rho^Q_G(\rb,t)=\rho^Q_G(\rb)\Theta(t),
\end{equation}
where $\Theta(t)$ is the Heaviside step function, $\Theta(t)=0$ for $t\le0$ and $\Theta(t)=1$ for $t>0$.
Under this form of the solute charge density, Eq.~\ref{eq:general} becomes
\begin{equation}
\rho^q(\rb,t)=-\rho_G^Q(\rb)\brac{\Theta(t) - \int_0^\infty ds \Theta(t-s) \mathcal{L}^{-1}\curly{\frac{1}{\varepsilon(\omega)}}}.
\label{eq:gen2}
\end{equation}
In order to further characterize the collective dynamical response of a polar solvent
to the solute charge distribution, we now need to assume an analytic form for $\varepsilon(\omega)$.
In general, $\varepsilon(\omega)$ exhibits many timescales encompassing the various types of motion
in a polar solvent.
Low frequency modes typically govern the long-time, collective polarization response that is responsible
for dielectric screening of the solute charge.
Here, we are concerned mainly with this low frequency dynamical response and the longer-time solvent
relaxation.
High frequency modes include electronic response (absent in point charge models), inertial and
librational motions, and molecular vibrations,
as well as any faster collective polarization fluctuations~\cite{Maroncelli:Nature,Ladanyi:JPC:1995,Stratt:JPC:1996}.
In this work, we coarse-grain out these high frequency modes and use a description of $\varepsilon(\omega)$
involving a single-Debye relaxation process~\cite{debye1929polar}
\begin{equation}
\varepsilon(\omega)=\varepsilon_h + \frac{\varepsilon - \varepsilon_h}{1-i\omega \tau_{\rm D}},
\label{eq:debye}
\end{equation}
where $\tau_{\rm D}\approx8$~ps is the Debye relaxation time~\cite{van1998systematic},
which can be obtained from fitting $\varepsilon(\omega)$ or through approximate statistical mechanical models
involving rotational diffusion~\cite{toda1991statistical}.
The static dielectric constant of SPC/E water is $\varepsilon\approx72$~\cite{van1998systematic}.
This form of $\varepsilon(\omega)$ arises from introducing a frequency cutoff, $\omega_h$,
such that processes above this frequency are assumed to occur effectively instantaneously.
At the frequency $\omega_h$, the Debye process, represented by the second term in Eq.~\ref{eq:debye},
is no longer the dominant relaxation channel, and $\varepsilon(\omega)\approx \varepsilon_h$ for $\omega>\omega_h$.
Fitting $\varepsilon(\omega)$ to Debye forms yields
$\varepsilon_h \approx 4$~\cite{Neuman:JCP:1986,Kindt,BENEDUCI200855,Schroder:JPCA:2015,C7CP02884A}.
%
%
Higher frequency motions, such as additional Debye-like relaxation processes
and short-time intertial Gaussian decays~\cite{Neuman:JCP:1986,Ladanyi:JPC:1995,Stratt:JPC:1996,Kindt,Maroncelli:Nature,BENEDUCI200855,Schroder:JPCA:2015,C7CP02884A},
are well understood and can be readily incorporated into more complex forms of $\varepsilon(\omega)$,
but this is beyond the scope of the current work.

The inverse Laplace transform of $\varepsilon^{-1}(\omega)$ can now be evaluated to yield
\begin{equation}
\mathcal{L}^{-1}\curly{\frac{1}{\varepsilon(\omega)}} = \frac{\varepsilon_h - \varepsilon}{\tau_{\rm D} \varepsilon_h^2} e^{-t/\tau_{\rm L}} + \frac{\delta(t)}{\varepsilon_h},
\label{eq:laplace}
\end{equation}
where $\tau_{\rm L}$ is the solvent longitudinal relaxation time defined by
\begin{equation}
\tau_{\rm L}=\left(\frac{\varepsilon_h}{\varepsilon}\right)\tau_{\rm D}.
\label{eq:longitudinal}
\end{equation}
Finally, upon insertion of Eq.~\ref{eq:laplace} into Eq.~\ref{eq:gen2}, we arrive at an expression for the time-dependent
induced solvent charge density,
\begin{equation}
\rho^q(\rb,t)=-\para{1-\frac{1}{\varepsilon}} \rho_G^Q(\rb)\Theta(t) - \para{\frac{1}{\varepsilon}-\frac{1}{\varepsilon_h}} \rho^Q_G(\rb) e^{-t/\tau_{\rm L}} \Theta(t).
\label{eq:rhoqrt}
\end{equation}
The first term in Eq.~\ref{eq:rhoqrt} corresponds to the equilibrium limit of the solvent response;
the solute charge density is partially screened by the solvent with static dielectric constant $\varepsilon$.
The second term in Eq.~\ref{eq:rhoqrt} interpolates between an effective screening due to the coarse-grained fast-time response at $t=0_+$
and the equilibrium limit as $t\rightarrow\infty$ on a timescale of $\tau_{\rm L}$.
The thermodynamic evolution of the system is encoded in the time-dependent energy gap, $\Delta E(t)$,
where 
\begin{equation}
\Delta E(t) = E_1(t) - E_0(t) = \Psi_1(\Rbar(t))-\Psi_0(\Rbar(t))
\end{equation}
is the difference in Hamiltonians of the excited ($\lambda=1$) and ground ($\lambda=0$) states at time $t$,
and $\Psi_\lambda(\Rbar)$ is given by Eq.~\ref{eq:psi}.
For the process considered here, there is no solute prior to insertion, $\rho^Q_0(\rb)=0$,
such that $E_0(t)=\Psi_0(\Rbar(t))=0$, and
the energy gap is given by the solvation energy,
\begin{equation}
\Delta E(t) = \Psi_1(\Rbar(t)) = \int d\rb \int d\rb' \frac{\rho^Q_G(\rb',t)\rho^q(\rb;\Rbar(t))}{\left| \rb-\rb'\right|},
\end{equation}
Using the solvent charge density given by Eq.~\ref{eq:rhoqrt}, 
we obtain
\begin{equation}
\Delta E(t) = \Theta(t) \Delta E^c(Q;l,\varepsilon) + e^{-t/\tau_{\rm L}} \Theta(t) \brac{\Delta E^c(Q;l,\varepsilon_h) - \Delta E^c(Q;l,\varepsilon)},
\label{eq:tfe}
\end{equation}
where the solvation energy, $\Delta E^c(Q;l,\varepsilon)=2\Delta G^c(Q;l,\varepsilon)$, is equal to twice
the charging free energy given by Eq.~\ref{eq:gcfe} for a Gaussian charge distribution
or by Eq.~\ref{eq:G_freeEnergy} for a Gaussian dipole, with $Q$ replaced by $\mu$.
As with the charge density, the first term in Eq.~\ref{eq:tfe} corresponds to the equilibrium limit of the solvation energy.
The non-trivial time evolution of the energy is given by the second term, which exponentially interpolates between the energy change due to fast-time response, via $\varepsilon_h$, at $t=0_+$ and the static, equilibrium solvation energy on the scale of $\tau_{\rm L}$.
%

%
The time dependence of solvation dynamics is characterized experimentally by the normalized,
nonequilibrium time correlation function
\begin{equation}
S\left(t\right)=\frac{\Delta E\left(t\right)-\Delta E\left(\infty\right)}{\Delta E\left(0\right)-\Delta E\left(\infty\right)},
\label{eq:S_func}
\end{equation}
where $t=0$ is the instant when an ion or a dipole is created in the solute. 
The nonequilibrium response function $S(t)$ can be measured experimentally using ultrafast spectroscopy,
e.g. time-dependent Stokes shift experiments~\cite{Maroncelli:Nature,Stratt:JPC:1996,Thompson:2011ax,castner1987subpicosecond,FlemingReview,BagchiBook}.
Thus, $S(t)$ offers the opportunity to readily bridge theory and experiment to interpret dynamic solvent response at the molecular scale.
Within our theory, this normalized response function can be obtained by inserting Eq.~\ref{eq:tfe}
into Eq.~\ref{eq:S_func} to yield
\begin{equation}
S\left(t\right)=1-\frac{1-\frac{1}{\varepsilon_h}}{1-\frac{1}{\varepsilon}}\Theta(t)+\frac{\frac{1}{\varepsilon_h}-\frac{1}{\varepsilon}}{1-\frac{1}{\varepsilon}}(e^{-\frac{t}{\tau_{\rm L}}}-1),\; t\geq 0. 
\label{eq:S_analytical}
\end{equation}
This form of $S(t)$ is equal to unity at $t=0$ and decays to zero at long times, as expected.
The second term in Eq.~\ref{eq:S_analytical} corresponds to an effectively instantaneous decay
due to high frequency ($\omega>\omega_h$) processes. 
The Debye relaxation is contained in the third term of Eq.~\ref{eq:S_analytical}, which is
a single exponential decay over a timescale of $\tau_{\rm L}$.
Note that this form of $S(t)$ is independent of the nature of the Gaussian charge distribution, e.g.
$S(t)$ is predicted to be the same for a Gaussian charge and a Gaussian dipole. 
%

%
\begin{figure*}[tb]
\begin{center}
\includegraphics[width=0.49\textwidth]{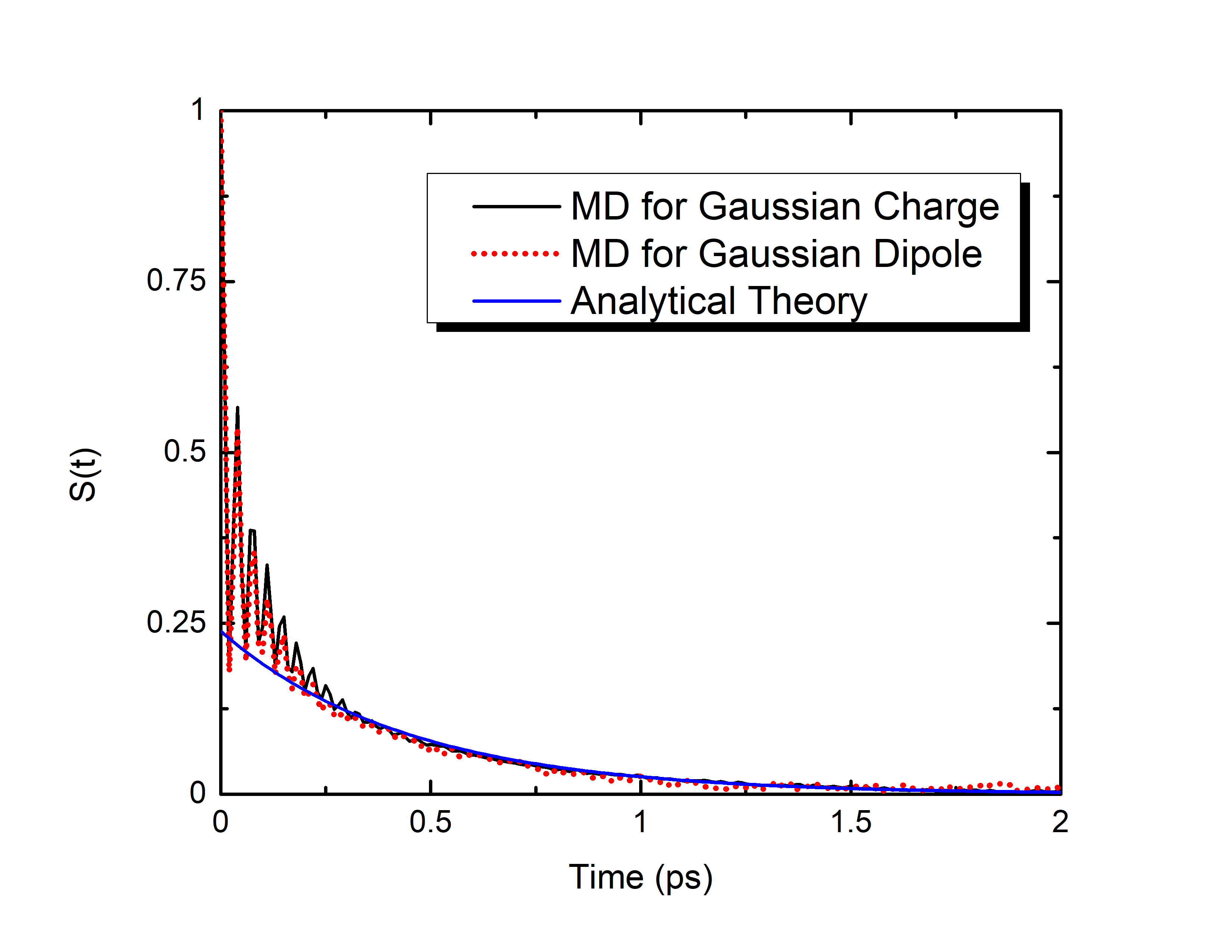}
\includegraphics[width=0.49\textwidth]{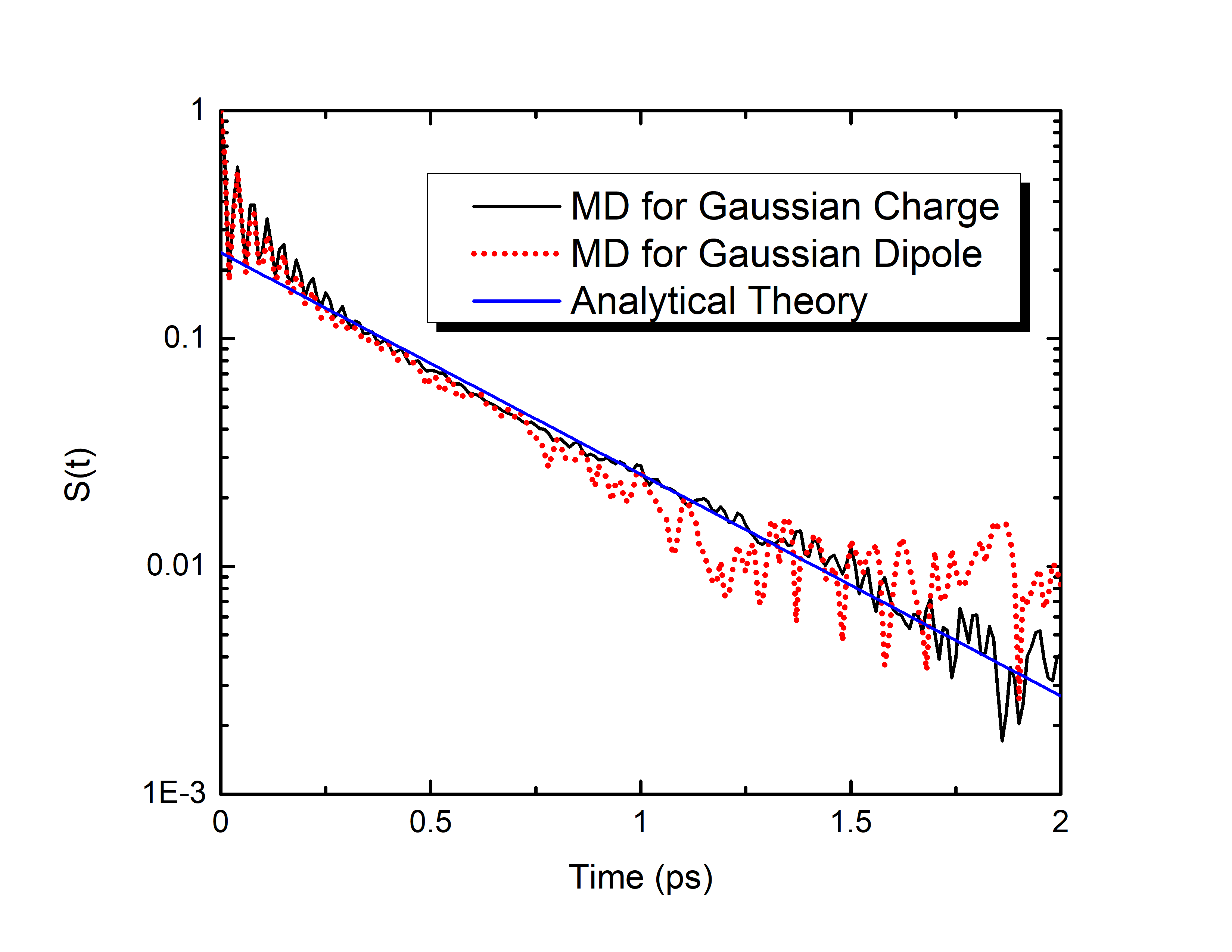}
\end{center}
\caption[]
{
Comparison between $S(t)$ computed from MD simulations (solid line) for a Gaussian charge with $Q=3e_0$ and $l=5.6$~\AA \ and $S(t)$ described by Eq.~\ref{eq:S_analytical} (dashed line) with parameters specified in the text. The exponential curve whose magnitude and rate are determined from the analytical theory can be considered as the asymptote of the long-time decay in the simulation result. Both plots contain the same data, the right panel is plotted on a semi-log scale.
}
\label{fig:Fig3New}
\end{figure*}

%
We performed molecular dynamics (MD) simulations of solvation dynamics in the SPC/E model of water by
turning on either a Gaussian charge or a Gaussian dipole at $t=0$ and computed
$S(t)$ as a nonequilibrium average over many simulation trajectories following Eq.~\ref{eq:S_func}.
%
The resulting nonequilibrium TCFs are shown in Fig.~\ref{fig:Fig3New} (curves labeled MD).
First, one can see that the two simulated $S(t)$ curves essentially overlap, showing
that the response to a Gaussian charge and Gaussian dipole are essentially the same, in
agreement with expectations from the above theory.
At very short times, $S(t)$ rapidly decays from one to about 0.25.
This fast decay is known to follow a Gaussian in time and arises from inertial motion
in response to instantaneous turning on of the solute field~\cite{Ladanyi:JPC:1995,Stratt:JPC:1996}.
Following this initial response, $S(t)$ exhibits damped oscillations followed by an exponential decay.
The coherent oscillations arise from underdamped, collective solvent motions on intermediate timescales,
wherein solvent molecules initially overpolarize due to the rapid switching on of the solvent
and then overcorrect, with this continuing until the oscillations are damped on a scale of about 0.5~ps~\cite{Stratt:JPC:1996}.
This damped oscillatory decay as well as the initial Gaussian decay process are high frequency
motions ($\omega>\omega_h$) and are therefore neglected in the theory for $S(t)$ in Eq.~\ref{eq:S_analytical}.
We also compare the prediction of Eq.~\ref{eq:S_analytical} with the simulation results
in Fig.~\ref{fig:Fig3New}.
The instantaneous drop in $S(t)$ at $t=0_+$ represented by the second term in Eq.~\ref{eq:S_analytical}
approximates the effects of all responses with frequencies higher than $\omega_h$.
This instantaneous drop is then followed by an exponential decay, described by the third term in Eq.~\ref{eq:S_analytical}.
The exponential decay captures the lower envelope of the response at times less than 0.5~ps,
indicating that the magnitude of the initial drop in $S(t)$ in adequately described by $\varepsilon_h$.
At longer times, Eq.~\ref{eq:S_analytical} accurately captures the long-time decay to equilibrium,
with the analytic prediction overlapping with the simulated results, for both the Gaussian charge and dipole.
This is emphasized by the examination of $S(t)$ on a logarithmic scale,
highlighting the accuracy of the long-time solvent response predicted by Eq.~\ref{eq:S_analytical}.
%

%
%

\begin{figure*}[tb]
\begin{center}
\includegraphics[width=0.48\textwidth]{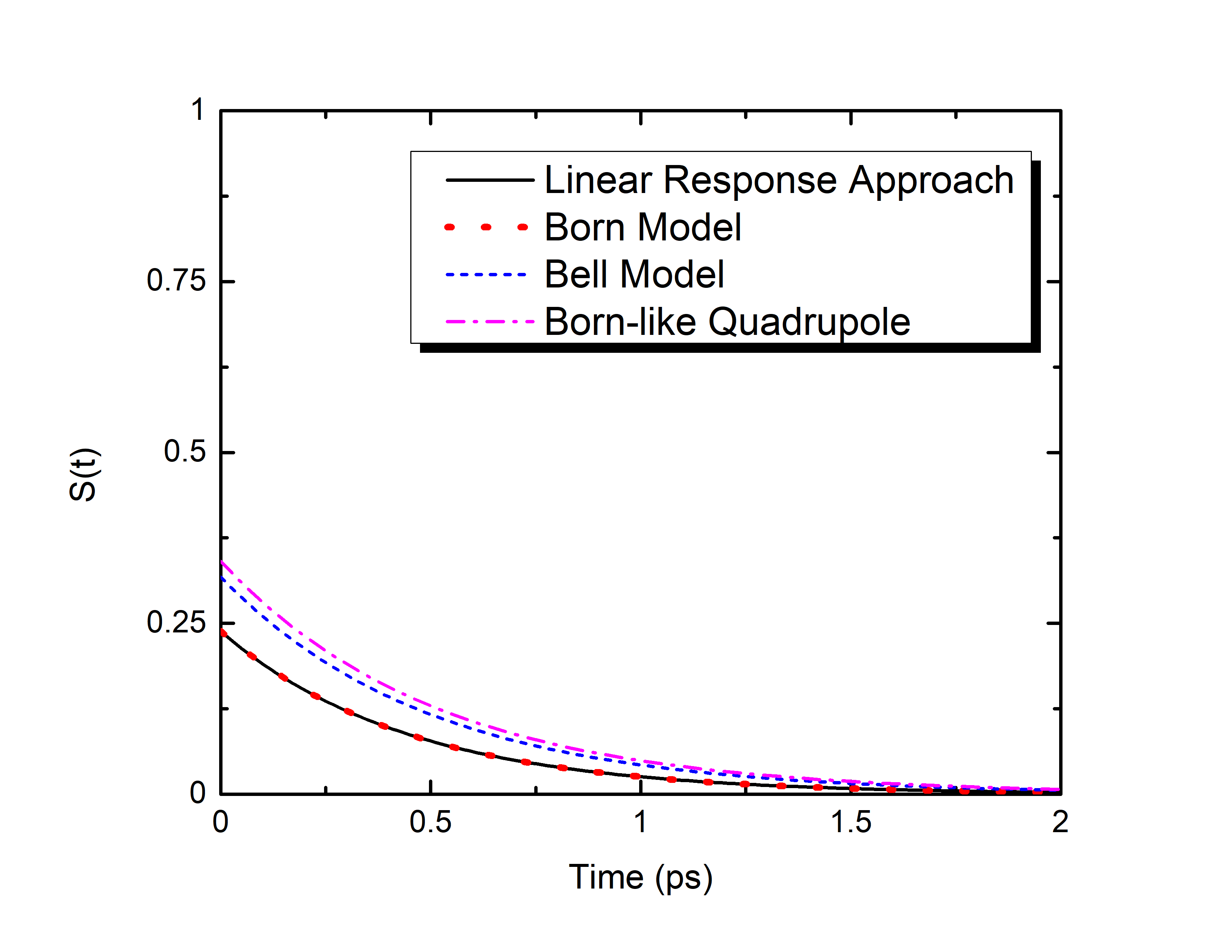}
\includegraphics[width=0.48\textwidth]{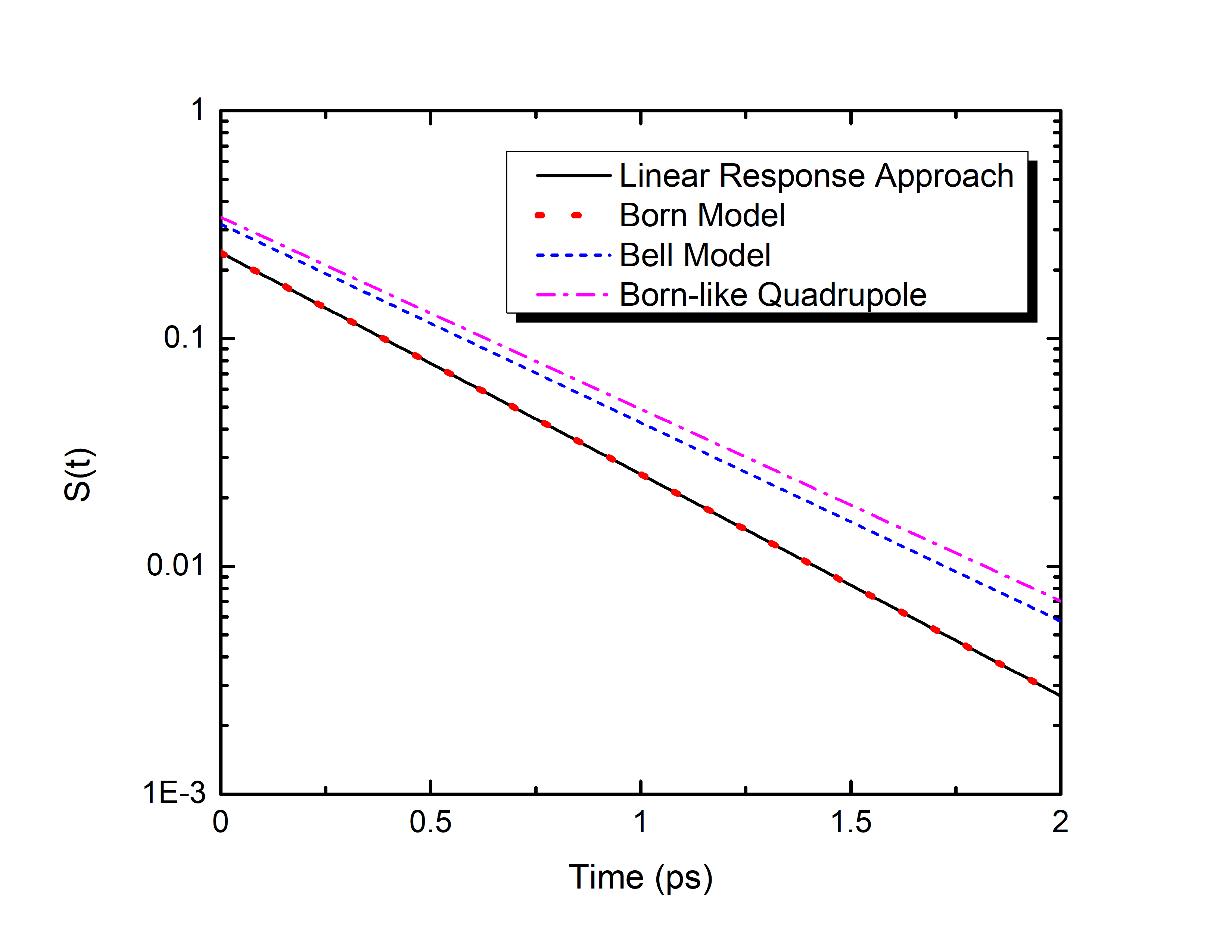}
\end{center}
\caption[analytical models]
{
Analytical predictions derived from our linear response theory-based model and the traditional models with hard cores.
The same parameters for the single Debye process are used in all the models.
Both plots contain the same data, the right panel is plotted on a semi-log scale.
}
\label{fig:models}
\end{figure*}

%
We emphasize that our theory for $S(t)$, Eq.~\ref{eq:S_analytical},
generally applies to any time-dependent slowly varying charge distribution, like the Gaussian distributions described here.
In these models, solute charges vary slowly over molecular length scales and induce local and linear responses in the solvent, such that Eq.~\ref{eq:linear_omega} is valid.
For example, Eq.~\ref{eq:S_analytical} correctly predicts the same time constant $\tau_L$ for the solvation dynamics
of both ionic and dipolar Gaussian-smoothed models, as shown in Fig.~\ref{fig:Fig3New}.
This longest time scale for the decay of $S(t)$ is the same as that derived with the classic Born model for ionic solvation.
However, the Bell model leads to a different time constant for solvation dynamics in response to a dipolar solute
composed of a hard core surrounding a point dipole,
\begin{equation}
\tau_L^{\rm Bell}=\left(\frac{{2\varepsilon}_h+1}{{2\varepsilon}+1}\right)\tau_D.
\label{eq:tl}
\end{equation}
The difference between $\tau_L$ and $\tau_L^{\rm Bell}$ originates from the different $\varepsilon$ dependences
of the free energy discussed in Sec. 2.
In general, the electrostatic boundary presented by the solute excluded volume leads to multipole-dependent
prefactors in Eq.~\ref{eq:tl}.
Moreover, the relative magnitudes of the short- and long-time responses predicted by hard-core models also differ depending on the order of the multipole, as shown in Fig.~\ref{fig:models}.
Therefore, hard-core models lead to non-negligibly different behaviors for the solvation of different multipoles~\cite{song1996gaussian},
such that the solvation dynamics involving solutes with hard cores
do not involve only the intrinsic dielectric response of the solvent, but involve a solute dependence as well.

\section{Conclusions}
Through the examination of the static and dynamic solvation of Gaussian charges and dipoles in a model of liquid water,
we have demonstrated that the intrinsic response of water to slowly-varying electrostatic perturbations is well-described by linear response theory.
The use of Gaussian-smoothed charge distributions removes the singularity
associated with point charge models at short distances,
eliminating the need for an excluded volume solute core.
The presence of such cores has obscured the interpretation of solvent response because the harshly repulsive cores inherently induce non-linear and asymmetric
thermodynamic and structural response in polar solvents~\cite{Remsing:JPCB:2016}. 
By avoiding these nonlinearities, we are able to probe and develop theories for the intrinsic dielectric response of water in the linear regime.
In classic dielectric continuum treatments, the introduction of a solute core creates a boundary
condition that can lead to free energies and relaxation times
that contain complicated, multipole-dependent functions of the dielectric constant.
In contrast, the solvent response to boundary-less Gaussian multipolar distributions
take on the same dielectric screening form.
This leads to multipole-independent timescales for solvation dynamics,
as well as the same solvent screening contribution to the solvation free energy,
i.e. the free energy is a product of a solvent term (involving $\varepsilon$) and a solute term
(involving the magnitude of the solute multipole moment). 
The intrinsic solvent response to electrostatic solute perturbations therefore does not depend
on the nature of the solute charge distribution itself beyond its magnitude and smearing length.
%

\section{Simulation Details}
Our MD simulations are performed with the LAMMPS software package~\cite{LAMMPS},
using the SPC/E water model~\cite{SPCE} and 8000 water molecules.
Lennard-Jones interactions are cutoff at 9.8~\AA \ and long-range electrostatic interactions
are evaluated using Ewald summation~\cite{CompSimLiqs} with a precision of ${10}^{-6}$.
The package is modified to include the potentials of a Gaussian charge and a Gaussian dipole, and the charging free energy is computed by direct summation following previous work~\cite{Remsing:JPCB:2016,Remsing2019}.
The system evolves in the NPT ensemble realized by a Nos\'{e}-Hoover barostat and thermostat with a timestep of 1.0~fs,
such that the temperature and pressure are maintained at $T = 300$ K and $P = 1$ atm~\cite{parrinello1981polymorphic,nose1983constant,nose1984unified,martyna1994constant,shinoda2004rapid}.
The equilibrium free energy is averaged from 9 trajectories, each monitored for 506~ps following a 20~ps equilibration between the solute charge and the pre-equilibrated bulk water.
The nonequilibrium response function is averaged from 900 trajectories for the Gaussian charge and 7200 trajectories for the Gaussian dipole.
%

\bibliographystyle{spphys}
\bibliography{GSD}


\end{document}